\documentclass[twoside,10pt]{article}

\usepackage{xcolor}
\usepackage{hyperref}
\hypersetup{colorlinks=true,linkcolor=violet,citecolor=violet,urlcolor=violet}

\usepackage{jmlr2e}

\usepackage{graphicx}
\usepackage{amsmath}
\usepackage{cleveref}
\usepackage{mathtools}
\usepackage{xspace}
\usepackage{multicol}
\usepackage{multirow}
\usepackage{booktabs}
\usepackage{caption}
\usepackage{subcaption}
\usepackage{enumitem}

\newcommand{\xmodel}{{GREENer}}

\jmlrheading{1}{2000}{1-48}{4/00}{10/00}{Peng Wang, Renqin Cai and Hongning Wang}

\ShortHeadings{Graph-based Extractive Explainer for Recommendations}{Wang, Cai and Wang}
\firstpageno{1}

\begin{document}

\title{Graph-based Extractive Explainer for Recommendations}

\author{\name Peng Wang\thanks{Equal contribution} \email pw7nc@virginia.edu \\
	\addr Department of Computer Science\\
	University of Virginia\\
	Charlottesville, VA 22903, USA
	\AND
	\name Renqin Cai\footnotemark[1] \email rc7ne@virginia.edu\\
	\addr Department of Computer Science\\
	University of Virginia\\
	Charlottesville, VA 22903, USA
	\AND
	\name Hongning Wang \email hw5x@virginia.edu\\
	\addr Department of Computer Science\\
	University of Virginia\\
	Charlottesville, VA 22903, USA
}

\maketitle

\begin{abstract}
Explanations in a recommender system assist users in making informed decisions among a set of recommended items. Great research attention has been devoted to generating natural language explanations to depict how the recommendations are generated and why the users should pay attention to them. However, due to different limitations of those solutions, e.g., template-based or generation-based, it is hard to make the explanations easily perceivable, reliable and personalized at the same time.

In this work, we develop a graph attentive neural network model that seamlessly integrates user, item, attributes, and sentences for extraction-based explanation. The attributes of items are selected as the intermediary to facilitate message passing for user-item specific evaluation of sentence relevance. And to balance individual sentence relevance, overall attribute coverage, and content redundancy, we solve an integer linear programming problem to make the final selection of sentences.  Extensive empirical evaluations against a set of state-of-the-art baseline methods on two benchmark review datasets demonstrated the generation quality of the proposed solution.
\end{abstract}
\maketitle


\newcommand{\tabincell}[2]{\begin{tabular}{@{}#1@{}}#2\end{tabular}}
\section{Introduction}

Nowadays, recommendations in online information service platforms, from e-commerce (such as Amazon and eBay) to streaming services (such as Netflix and youtube), have greatly shaped everyone's life, by affecting who sees what and when \citep{aggarwal2016recommender,fleder2009blockbuster,milano2020recommender}. 
%However, explanations about why a particular person gets a certain recommendation are not provided. 
Therefore, besides improving the quality of recommendations, explaining the recommendations to the end users, e.g., how the recommendations are generated \citep{tao2019the,wang2019explainable,ribeiro2016should} and why they should pay attention to the recommended content \citep{wang2018explainable,zhang2014explicit,xian2021ex3}, is also critical to improve users' engagement and trust in the systems \citep{bilgic2005explaining, herlocker2000explaining, sinha2002role}.

\begin{table*}[t]
    \centering
    \caption{Example explanations produced by 3 different types of explainable recommendation models.}
    \vspace{-2mm}
    \label{table:motive_example}
    \begin{tabular}{|c|p{6.2cm}|p{6.2cm}|}
        \hline
        Hotel & { Kimpton Hotel Eventi, NYC} & New Orleans Marriott \\
        \hline
        EFM & You might be interested in staff/room, on which this hotel performs well. & You might be interested in staff/location, on which this hotel performs well. \\
        \hline
        SAER & The room was spacious and the room was clean. I was very pleased with the staff, the hotel and staff were very friendly. & The location was great, the hotel was very nice, and the rooms were clean and comfortable. The room was spacious and the beds were extremely comfortable. \\
        \hline
        \xmodel{} & The view of the empire state building was incredible! The hotel was beautifully decorated, and the staff was very helpful. The room was huge and very clean, the bed comfy and the jacuzzi wonderful.  & There was a great view of the Mississippi river. Can't beat the location, just a few blocks from the heart of bourbon street. \\
        \hline
    \end{tabular}
    \vspace{-2mm}
\end{table*}

%Providing explanations of recommendations is attracting more and more attention for the purpose of making recommendations more understandable and reliable. 
To be helpful in users' decision making, the system-provided explanations have to be easily perceivable, reliable, and personalized.
%Natural language based explanation methods have been one of the mainstream solutions in the community. Comparing to 
Template-based explanations have been the dominating choice, where various solutions were developed to extract attribute keywords or attribute-opinion pairs from user reviews \citep{xian2021ex3,wang2018explainable,zhang2014explicit,tao2019the} or from pre-existing knowledge graph \citep{xian2019reinforcement,wang2019explainable} to form the explanation content about a specific item for a target user. The fidelity of the generated explanations can be improved by careful quality control in the algorithms' input.
But the predefined templates lack desirable diversity, and their rigid format and robotic style are less appealing to ordinary users \citep{yang2021explanation,li2020generate}. 

On the other hand, due to the encouraging expressiveness of the content generated from neural language models \citep{duvsek2020evaluating,radford2018improving,brown2020language}, an increasing number of solutions adopt generative models for explanation generation \citep{yang2021explanation,li2020generate,li2017neural}. The generated content from such solutions are generally believed to have better readability and variability. %Natural language based explainable methods are to produce a text snippet to describe which attributes of the item attract the user. 
Nevertheless, the high complexity of neural language models prevents fine-grained control in its generated content. And the success of such models heavily depends on the availability of large-scale training data. Due to the lack of observations about opinionated content from individual users on specific items, the generation from such models can hardly be personalized. On the contrary, it has been observed that such models' output tend to be generic and sometimes even less relevant to target items \citep{yang2021explanation}. 

To understand the aforementioned advantages and limitations of these two types of explanation generation methods, we extract a few sample outputs from one typical solution of each type trained on the same hotel recommendation dataset in Table \Cref{table:motive_example}. We chose Explicit Factor Model (EFM) \citep{zhang2014explicit} to represent template-based solutions, and Sentiment Aligned Explainable Recommendation (SAER) \citep{yang2021explanation} to represent neural generative solutions. The robotic style of EFM's explanation content can be easily recognized, e.g., only the attribute keyword changes across its output for different items. Even if the recommended hotels in the example are indeed featured with \emph{staff} or \emph{location}, such generic content hurts the trustworthiness of the explanations. On the other hand, although SAER's output style is more diverse, its content is quite generic; especially when comparing across items, the aspects mentioned are less specific about the target items. This is also problematic when a user needs to choose from a set of recommended items based on their explanations.

To make the explanations easily perceivable, reliable,  and also personalized, we propose an extractive solution, named GRaph Extractive ExplaiNer (\xmodel{}), to extract sentences from existing reviews for each user-item pair as explanations. By collectively selecting from existing review content, the extracted sentences maintain the readability from human-written content, and thus make the explanations easily perceivable. 
For a given pair of user and item, the past reviews from the user suggest his/her preferences on different aspects/attributes of this type of items; and the past reviews describing the item suggest its commonly discussed aspects. Hence, specificity about the user and item can be captured, which leads to personalized explanations. 
%Sentences from these reviews provide sources to produce explanations personalized to the pair of user and item. 
And because of the aggregation among user-provided content about the item, the reliability of the selected content can also be improved.

Accurate extraction from existing content as explanations is however non-trivial. First, not all the sentences in a user review are relevant to the item. For example, it is very common to encounter users' personal experiences mentioned in a review. Such content is clearly unqualified as explanations and should be filtered. 
Second, the user and item should play different roles in selecting the sentences for explanation: the item suggests the set of relevant aspects, while the user suggests where the attention should be paid to. Therefore, the interplay between the user and item should be carefully calibrated when evaluating a sentence's relevance.
Third, the selected sentences should cover distinct aspects of an item; and it is apparently undesirable to repeatedly mention the same aspect in different sentences when explaining an item. However, it is expected that an item's popular attributes will be mentioned in multiple users' reviews with some content variations. Avoiding such nearly duplicated content becomes necessary and challenging.

To address these challenges in extractive explanation generation, we develop a graph attentive neural network model that seamlessly integrates user, item, attributes and sentences for sentence selection. For a collection of items, we first extract frequently mentioned attributes as the intermediary to connect users and items with sentences, i.e., the connectivity on the graph suggests who mentioned what about the item. As a result, sentences not related to any selected attributes are automatically filtered. %, which are clearly not qualified as explanations. 
To handle data sparsity when estimating the model parameters, we employ pre-trained language models \citep{devlin2018bert} for attribute words' and sentences' initial encoding. 
Through attentive message passing on the graph, heterogeneous information from user, item and attributes about the candidate sentences is aggregated for user-item specific evaluation of sentence relevance. 
However, because each sentence is independently evaluated by the neural network, content redundancy across sentences cannot be directly handled. We introduce a post-processing strategy based on Integer Linear Programming to select the final top-K output, where the trade-off between relevance and redundancy is optimized.

To investigate the effectiveness of \xmodel{} for explanation generation, we performed extensive experiments on two large public review datasets, Ratebeer \citep{yang2021explanation} and TripAdvisor \citep{wang2010latent}. Compared with state-of-the-art solutions for explanations, \xmodel{} improved the explanation quality in both BLEU \citep{papineni2002bleu} and ROUGE \citep{lin2004rouge} metrics. Our ablation analysis further demonstrated the importance of modeling these four types of information source for explanation generation, and also the importance of a graph structure for capturing the inter-dependency among them. Our case studies suggest that our produced explanations are more perceivable, specific to the target user-item pair, and thus more reliable.

\section{Related Work}
Numerous studies have demonstrated that explanations play an important role in helping users evaluate results from a recommender system \citep{swearingen2001beyond,symeonidis2008providing,zhang2018explainable}.
And various forms of explanations have been proposed, from social explanations such as ``\textit{X, Y and 2 other friends like this.}'' \citep{sharma2013social}, to item relation explanations such as ``\textit{A and B are usually purchased together.}'' \citep{ma2019jointly,wang2019explainable,xian2019reinforcement}, and opinionated text explanations such as ``\textit{This \textbf{phone} is featured with its \textit{high-resolution} \textbf{screen}.}'' \citep{zhang2014explicit,wang2018explainable,yang2021explanation}, which is the focus of this work.

There are currently three mainstream solutions to generate opinionated textual explanations, namely template-based, generation-based, and extraction-based methods. They all work on user-provided item reviews to create textual explanations.
In particular, template-based methods predict important attributes of the target item together with the sentiment keywords from user reviews to fill in the slots in those manually crafted templates. As typical solutions of this type, EFM \citep{zhang2014explicit} and MTER \citep{wang2018explainable} predict important item attributes and corresponding user opinion words for a given recommendation via matrix factorization and tensor factorization. EX$^3$ extracts key attributes to explain a set of recommendations, based on the idea that the selected attributes should predict users' purchasing behavior of those items \citep{xian2021ex3}. CountER employs counterfactual reasoning to select the important aspects for explanation \citep{tan2021counterfactual}. The main focus in these template-based methods has been devoted to identify the most important item attributes and user opinion, i.e., to improve reliability and personalization; but its lack of content variability and robotic explanation style make such explanations less appealing to the end users. 

To increase content diversity in the provided explanations, neural language models are employed in generation-based methods to synthesize natural language explanations. As an earlier work, NRT models explanation generation and item recommendation in a shared user-item embedding space, where its predicted recommendation rating is used as part of the initial state for corresponding explanation generation \citep{li2017neural}. NETE shared a very similar idea with NRT, but it further confines the generation to cover specific item attributes that are selected by a separated prediction module \citep{li2020generate}. SAER constrains the sentiment conveyed in the generated explanation content to be close to the item's recommendation score \citep{yang2021explanation}.  
However, due to the high complexity of neural language models, it is very hard to control such models' content generation at a fine granularity. As a result, the reliability of its generation is questionable. Furthermore, such methods tend to generate generic content to fit the overall data distribution in a dataset. Hence, on the user side, the explanation is less personalized; and on the item side, the explanation could be even less relevant (e.g., overly generic). 

%Graph attention networks (GATs) have achieved great success in many tasks, like text classification or reading comprehension~\cite{velivckovic2017graph, wang2019heterogeneous, hu2020heterogeneous, wang2020heterogeneous}. GATs consider the graph structure among data and model the high-order and complex relations among nodes. With the message passing, GATs learn node representations by aggregating information from nodes' neighboring nodes. Based on the encoded node representations, GATs make downstream task predictions. Early applications of GATs focus on homogeneous graphs which consists of the same type of nodes , e.g., multi-class node classifications in Cora, Citeseer and Pubmed citation network datasets~\cite{velivckovic2017graph}. Recently, researchers have applied GATs to heterogeneous graphs with multiple types of nodes, e.g., text summarization~\cite{wang2020heterogeneous}. They have shown that GATs can capture multiple relations among multiple types of nodes. Inspired by these work, we adopt GATs for our task to capture multiple relations among user, item, feature and sentence nodes, i.e., the co-occurrence relations of user-feature and item-feature, and the semantic relations of feature-sentence.

Extraction-based solutions directly select representative sentences from the target item's existing reviews. And our proposed solution falls into this category. NARRE selects the most attentive reviews as the explanation, based on the attention originally learned to enrich the user and item representations for recommendation  \citep{chen2018neural}. CARP uses the capsule network for the same purpose \citep{li2019capsule}. \citet{wang2018reinforcement} adopt reinforcement learning to extract the most relevant review text that matches a given recommender system's rating prediction. 
In nature, extraction-based solutions model the affinity between user-item pairs with sentences, which is very sparse: a user review is typically short and a user typically does not write many reviews. Personalized explanation is hard to achieve in such a scenario. Our solution breaks this limitation by introducing item attributes as an intermediary, which not only alleviate sparsity issue but also improves specificity and reliability of the generated explanations (e.g., the sentences will only cover attributes associated with the target item). We should note that the extraction-based solutions are restricted to an item's existing reviews; for items with limited exposure, e.g., a new item, it is hard to generate informative explanations in general. A very recent work combines extractive and generative methods for explanations \citep{yang2021comparative}, which has potential to solve the challenge. We leave this direction as our future work.  
%However, these extractive solutions still depend on the co-occurrence of user-item-sentence, which is sparse. Different from these solutions, \xmodel{} makes use of the co-occurrence of user-feature and item-feature which are denser than that of user-sentence and item-sentence. This denser co-occurrence allows \xmodel{} to select sentences that are more relevant to the pairs of users and items. 

% add graph neural network discussion

% because extraction-based solutions are restricted to an item's existing reviews, which are subject to the availability and quality of existing content. For items with limited exposure, e.g., a new item, such solutions can hardly provide any informative explanations.  

\section{Graph Extractive Explainer for Recommendations}
In this section, we describe our proposed extractive explanation solution \xmodel{} in detail. At the core of \xmodel{} is a graph neural network, which integrates heterogeneous information about a user, a recommended item, and all candidate sentences via attentive message passing. To alleviate the observation sparsity issue at the user-item level, we introduce item attributes as an intermediary to connect user, item and sentences. Finally, the extraction is performed by solving an integer linear programming problem to balance individual sentence relevance, overall coverage, and content diversity in the final selections.  

\begin{figure*}[t]
\centering
\includegraphics[width=0.95\textwidth]{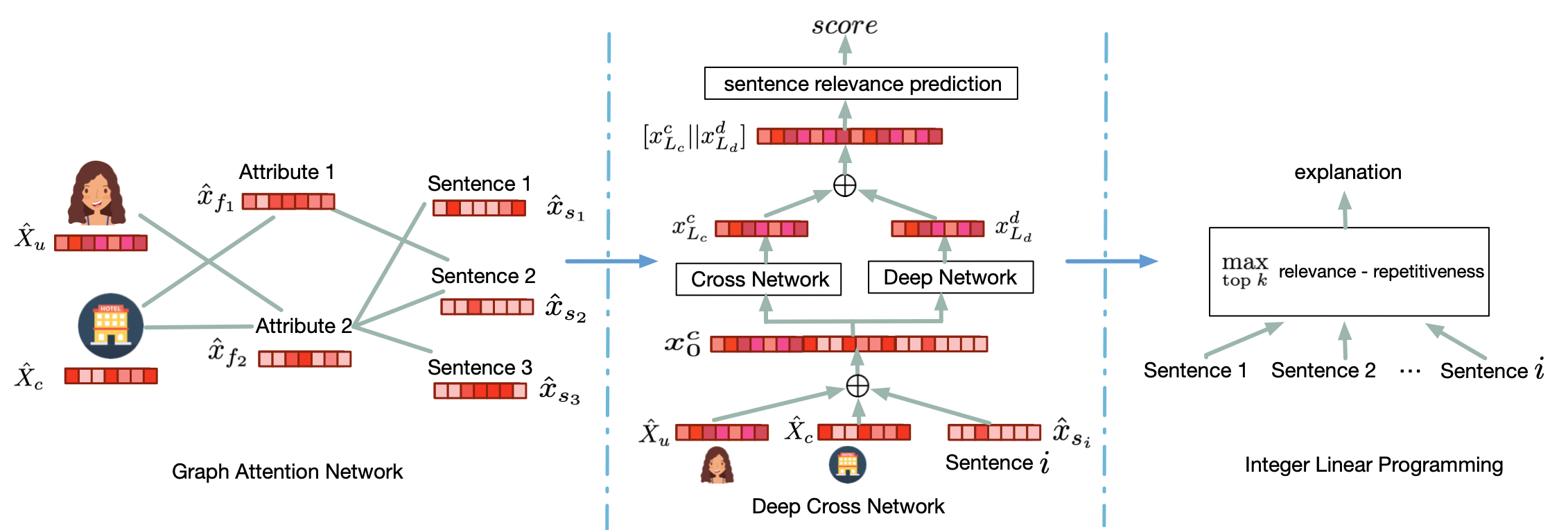}
\vspace{-2mm}
\caption{Illustration of GRaph Extractive ExplaiNer (GREENer). For a pair of user and item, \xmodel{} utilizes graph attention network and deep cross network to encode past sentences written by the user and past sentences describing the item. Then it utilizes Integer Linear Programming to select sentences as an explanation.}
\label{fig:xmodel_design}
\vspace{-2mm}
\end{figure*}

\subsection{Problem Setup \& Notations}
We first define the notations employed in this paper.
Denote the candidate recommendation item set as $\mathcal{C}$, the set of users as $\mathcal{U}$, and the vocabulary of text reviews as $\mathcal{V}$. We use $|\cdot|$ to denote the cardinality of a set.  
%We assume there are $G$ reviews $Re$ written by users $\mathcal{U}$ towards items $\mathcal{C}$. Sentences in these reviews are denoted as $\mathcal{D}=\{ \mathcal{S}_{uc}\}_{u\in \mathcal{U}, c \in \mathcal{C}}$. 
% We collect all the reviews from $\mathcal{U}$ about $\mathcal{C}$ and segment them into sentences. Then we denote $\mathcal{S}_{uc}=\{s_{uc}^i\}_{i=1}^N$ as all the sentences from user $u$ about item $c$, where each sentence $s_{uc}=\{w_i\}_{i=1}^{T}$ consists of $T$ words $w \in \mathcal{V}$. We aggregate sentences from user $u$ over all items into a set $\mathcal{S}_{u}=\{\mathcal{S}_{uc}\}_{c\in \mathcal{C}}$; and similarly, we aggregate sentences about item $c$ from all users into $\mathcal{S}_{c}=\{\mathcal{S}_{uc}\}_{u\in \mathcal{U}}$. The attributes $\mathcal{F}$ are items' popular properties mentioned in the whole review corpus, which are a subset of words in $\mathcal{V}$. For a pair of a user $u\in \mathcal{U}$ and an item $c\in \mathcal{C}$, an extractive explanation model is to select $K$ sentences $\{s^i\}_{i=1}^K$ from the union of sentences $\mathcal{S}_{u} \bigcup \mathcal{S}_{c}$, such that these $K$ sentences best describe how the item is relevant to the user's preference. 
We collect all the reviews from $\mathcal{U}$ about $\mathcal{C}$ and segment them into sentences. Then we denote $\mathcal{S}_{uc}=\{s_{uc}^i\}_{i=1}^N$ as all the sentences from user $u$ about item $c$, where each sentence $s_{uc}=\{w_i\}_{i=1}^{T}$ consists of $T$ words $w \in \mathcal{V}$. We aggregate sentences from user $u$ over all items into a set $\mathcal{S}_{u}=\{\mathcal{S}_{uc}\}_{c\in \mathcal{C}}$; and similarly, we aggregate sentences about item $c$ from all users into $\mathcal{S}_{c}=\{\mathcal{S}_{uc}\}_{u\in \mathcal{U}}$. The attributes $\mathcal{F}$ are items' popular properties mentioned in the whole review corpus, which are a subset of words in $\mathcal{V}$. For a pair of a user $u\in \mathcal{U}$ and an item $c\in \mathcal{C}$, an extractive explanation model is to select $K$ sentences $\{s^i\}_{i=1}^K$ from the union of sentences $\mathcal{S}_{u} \bigcup \mathcal{S}_{c}$, such that these $K$ sentences best describe how the item is relevant to the user's preference.

\subsection{Neural Graph Model for Sentence Encoding}
Measuring the relatedness between a candidate sentence and a target pair of user and item is vital for precise extractive explanation. 
On a given training corpus, the relatedness can be inferred based on the observed associations between user, item, and sentences in the reviews. For example, how does a user usually describe a type of items; and how is an item typically commented by a group of users. But such observations are expected to be sparse, as a user often provides only a handful of reviews. 

\xmodel{} addresses the sparsity issue by introducing item attributes as the intermediary between a user-item pair and associated sentences, and accordingly captures the relatedness from two complementary perspectives.
First, the observed sentences written by the user and those describing the item, connected via the attributes appearing in these sentences, suggest the direct dependence between a sentence and the pair of user and item. This co-occurrence relation forms a heterogeneous graph among users, items, attributes, and sentences, suggesting their contextualized dependency. \xmodel{} leverages an attentive graph neural network to model this direct dependence structure.
Second, the learnt representations of users, items and sentences also suggest the dependence in the embedding space. \xmodel{} utilizes the feature crossing technique to capture this indirect dependence. In the following, we provide the technical details of these two important components. 

\subsubsection{Attentive Graph Structure}
%Graph is a natural choice to model entities of  different types. 
\xmodel{} utilizes a heterogeneous graph to capture the contextualized dependence among users, items, attributes, and sentences. For each review written by a user about an item,
% For a pair of a user and an item,
we construct a graph consisting of the user, the item, all sentences written by the user, all sentences describing the item, and the attributes mentioned in these sentences. The detailed graph structure is illustrated in Figure~\ref{fig:xmodel_design}. Attributes serve as a bridge to connect the user, item and sentences, via the observed co-occurrences among these four types of entities.
Attentive message passing on the graph integrates information from these different types of entities to form their latent representations, which can be used to predict if a candidate sentence is relevant to a given user-item pair, i.e., extract as explanation. 
%To propagate the personalized choices of attributes to the selection of sentences, the attributes are connected to sentences with edges. 
%Feeding the graph into graph attention network, we obtain the sentence hidden representations and attribute hidden representations. Based on sentence hidden representations, we predict whether a sentence is included as an explanation or not. Likewise, based on attribute hidden representations, we predict whether an attribute is mentioned in the explanation or not. The joint supervision from these two sources are used to train the model. 

% The graph structure among the user, the item, the sentences and features allow \xmodel{} to leverage the co-occurrence of user-feature, item-feature and attribute-sentence jointly. 
Next, we will zoom into the detailed design of this attentive graph for learning the representations of users, items and sentences. 

\noindent\textbf{Nodes.} A graph $g$ is created for each item-user pair, which consists of four types of nodes: a user node $u$, an item node $c$, $|S_g|$ sentence nodes where $\mathcal{S}_g=\mathcal{S}_u \bigcup \mathcal{S}_c$,
%represents the union of sentences written by the user $u$ and sentences talking about the item $c$, 
and $M$ attribute nodes $\{f_g^i\}_{i=1}^M\; \text{where}\; M \leq |\mathcal{F}|$ represents the attributes appearing in sentences $\mathcal{S}_g$. Note the set of sentences $\mathcal{S}_{uc}=\{s_{uc}^i\}_{i=1}^N$ written by  user $u$ about item $c$ is a subset of $\mathcal{S}_g$. 
Empirically, sentences in $\mathcal{S}_{g}$ that do not contain any attribute specific to item $c$ can be filtered to further improve the accuracy of the finally selected sentences.

The input representation of the user node is a dense vector $X_u$, obtained by mapping the user index $u$ through the input user embeddings $E_u \in \mathbb{R}^{|U|\times d_{u}}$. Likewise, the input representation of the item node $X_c$ is obtained in the same manner from $E_c \in \mathbb{R}^{|\mathcal{C}|\times d_{c}}$ accordingly. To obtain good semantic representations, instead of learning sentence representations from scratch, we take advantage of the pre-trained language model BERT \citep{devlin2018bert} to encode sentences $\mathcal{S}_{g}$ into input node representations. Specifically, we fine-tuned BERT on the review text data in the training set. Then we feed sentences into the fine-tuned BERT to obtain their embedding vectors $\{X_{s_{g}^i}\}_{i=1}^{|\mathcal{S}_g|}$ as input representations of sentence nodes. Likewise, the input representation of attribute nodes $\{X_{f_{g}^i}\}_{i=1}^M$ is also pre-trained on the review text data using GloVe \citep{pennington2014glove}. 

\noindent\textbf{Edges.} To capture the co-occurrence among different entities in the observed review content, we introduce an edge $e_{uf}$ connecting user node $u$ to attribute node $f$ if the attribute was used by the user in his/her training reviews. Likewise, edge $e_{cf}$ is introduced to connect item node $c$ and attribute node $f$ if the attribute was used to describe the item. Finally, an edge $e_{fs}$ is introduced to connect attribute node $f$ to sentence node $s$ if the sentence contains the attribute word. Notice that all the edges are non-directional by design, as shown in Figure \ref{fig:xmodel_design}. As a result, the attributes serve as a bridge to connect the user and item with individual sentences. For example, a user can now be associated with sentences from other reviews about the item, and so can the item and sentences be. 
% Through the path containing edges $e_{fs_i} \rightarrow e_{fs_j}$, the sentence $s_i$ is connected to the sentence $s_j$. Consequently, the relationship between them are modeled. 
In this work, we only consider the binary edge weight; other type of edge weights, e.g., continuous weights, are left as our future work. 

\noindent\textbf{Attentive Aggregation Layer.}
Given a constructed graph $g$ with nodes $\{X_u, X_c, X_f, X_s\}$ and edges $\{e_{uf}, e_{cf}, e_{fs}\}$, we adopt the attention mechanism from the graph attention networks \citep{velivckovic2017graph} to encode co-occurrence information into node representations. Specifically, we stack $L$ graph attention layer to map the input node representations into the output node representations $\{\hat{X}_u, \hat{X}_c, \hat{X}_f, \hat{X}_s\}$. Due to the recursive nature of graph attention, we only use one layer as an example to illustrate the design in our solution. For example, in the $l$-th layer, the inputs to the graph attention layer are $H^l = \{H_u, H_c, H_f, H_s\}$, which correspond to hidden representations of user node, item node, attribute nodes and sentence nodes obtained from the $(l-1)$-th layer. For the $i$-th node $h^l_i$ in the graph, we obtain attention weights $\alpha^l$ for its connected nodes as, 
\begin{align*}
\alpha^l_{ij} &= \frac{\exp(z^l_{ij})}{\sum_{j'\in \mathcal{N}_i} \exp(z^l_{ij'})}
\\
z^l_{ij} &= \text{LeakyReLU}\big(W^l_a[W^l_qh^l_i||W^l_kh^l_j]\big) 
\end{align*}
where $\mathcal{N}_i$ refers to neighborhood of node $i$. $\{W^l_a, W^l_q, W^l_k\}$ are parameters to be estimated and $||$ denotes the concatenation operation. 

With the attention weights, we obtain the output hidden representation of node $i$ in the $l$-th layer as
% \begin{align}
%     h^{l+1}_i = \sigma (\sum_{j \in \mathcal{N}_i}(\alpha^l_{ij}h^l_i))
%     \label{eq:gat_merge}
% \end{align}
\begin{align}
    h^{l+1}_i = \sigma \Big(\sum_{j \in \mathcal{N}_i}\alpha^l_{ij}h^l_j\Big)
    \label{eq:gat_merge}
\end{align}

With the $d_h$ multi-head attention, we repeat the above process $d_h$ times and merge the output hidden representations from $d_h$ heads as the representation of node $i$ as $h^{l+1}_i = ||_{head=1}^{d_h} h^{l+1}_{head,i}$, where $h^{l+1}_{head,i}$ is obtained by Eq.~\eqref{eq:gat_merge}. 

Note that for the initial attention layer, we use the input node representations $\{X_u, X_c, X_f, X_s\}$ as the input $H^0$, and through $L$ attention layers, we use the output representations $H^{L}$ as the output node representations $\{\hat{X}_u, \hat{X}_c, \hat{X}_f, \hat{X}_s\}$.

\subsubsection{Feature Crossing}
The attentive graph structure captures the direct co-occurrence dependency between a user-item pair and sentences. From a distinct perspective, feature crossing models the indirect dependency among them on top of the graph representations. 
Following the design of Deep \& Cross network (DCN)~\citep{wang2017deep}, \xmodel{} applies feature crossing to model the representation-level interaction among $\hat{X}_u, \hat{X}_c$, and $\hat{X}_s$.
DCN is a combination of cross network and deep network in a parallel structure as shown in Figure~\ref{fig:xmodel_design}. The cross network is a stack of multiple cross layers, which can be written as
\begin{align*}
     x^{c}_{l+1} &= x^{c}_{0} x^{c\mathrm{T}}_{l} w^{c}_{l} + b^{c}_{l} + x^{c}_{l}
\end{align*}
where $x^{c}_{l}$ represents the hidden state in $l$-th layer of the cross network. The deep network is a multiple layer fully-connected neural network, which can be written as
\begin{align*}
    x^{d}_{l+1} &= f(W^{d}_l x^{d}_{l} + b^{d}_{l})
\end{align*}
where $x^{d}_{l}$ represents the hidden state in $l$-th layer of the deep network. $\{w^{c}_{l}, b^{c}_{l}, W^{d}_l, b^{d}_{l}\}$ are trainable parameters. These two networks take the same input, which is a concatenation of user, item and sentence output node representation from the group attention layers, as 
\begin{align*}
    x^{c}_{0} &= x^{d}_{0} = [\hat{X}_u || \hat{X}_c || \hat{x}_{s_i}]
\end{align*}

Through this DCN module, we can obtain the final output representation sentence $s$ in graph $g$ as 
\begin{align*}
    x^{cd}_s = [x^{c}_{L_{c}} || x^{d}_{L_{d}}]
\end{align*}
where $x^{c}_{L_{c}}$ and $x^{d}_{L_{d}}$ are the outputs from the cross network and deep network of  sentence $s$, respectively. It aggregates information about a sentence's relatedness to a user-item pair from their direct co-occurrence relation and indirect representation-level dependency. 

\subsection{Sentence Extraction}
Based on the encoded sentence representations, \xmodel{} learns to predict if the sentences are qualified explanations. In addition, \xmodel{} also predicts if an attribute should be mentioned in the extracted explanations, which forms a multi-task objective for parameter estimation. 
%Considering the task of predicting attributes appearing in explanations is related to the task of predicting sentences as explanations, we optimize these two tasks jointly, i.e., multi-task learning. 

\noindent\textbf{Multi-Task Objective.} 
In a given training corpus of review sentences, \xmodel{} is trained to rank sentences mostly related to the ground-truth sentences from user $u$'s review about item $c$ (i.e., $\mathcal{S}_{uc}$) above all other sentences. This is realized via a pairwise rank loss function. 
%Based on the observation that sentences that are not included in the ground-truth explanations can be also similar to sentences that are included in the ground-truth explanations, we use pairwise loss as the loss function of sentence predictions. Specifically, for each sentence, we obtain its similarities to every ground truth sentence and use the maximum similarity as its relevance to the ground truth. Based on sentences' relevance, we obtain their pairwise order. Then we obtain the unnormalized scores of sentences being selected. With the pairwise order and unnormalized scores, we obtain the pairwise loss. 
Specifically, the ranking score $g(s_i)$ of sentence $s_i$ is obtained by its feature crossing representation $x^{cd}_i$ via a linear mapping $g(s_i) = \langle W^s_o\,, x^{cd}_i \rangle$.

The relevance of a candidate sentence $s_i$ against $\mathcal{S}_{uc}$ can be simply realized via an indicator function, i.e., whether the $s_i$ is in $\mathcal{S}_{uc}$. To relax this, we choose to measure the similarity $r_{i}$ between $s_i$ and $\mathcal{S}_{uc}$,
\begin{align}
\label{eq-stn-sim}
    r_{i} = \max_{s^{j}_{uc} \in \mathcal{S}_{uc}} \textit{sim} (s_i, s^{j}_{uc})
\end{align}
where $\textit{sim}(\cdot)$ can be any text similarity metric that measures the semantic similarity between a pair of sentences. In our experiments, we used BLEU score, such that sentences in $\mathcal{S}_{uc}$ always have the highest similarity.

Under this similarity-based notion of sentence relevance, the pairwise ranking loss on a set of candidate sentences can be computed as follows,
\begin{align*}
    L_s = -\sum_{s_i, s_j\in S_g} \text{sign}(r_i-r_j) \log \; \sigma\big(g(s_i)-g(s_j)\big)
\end{align*}
where $\sigma(\cdot)$ is the sigmoid function. 
%The combination of sentence-wise similarity score and pairwise loss helps \xmodel{} to learn a larger ranking score $f(s_i)$ for sentence that is more similar to the ground-truth review.

% \begin{align*}
%     $\mathbf \textit{sign}(r_i-r_j)$  :=
%  \begin{cases}1~&{\text{ if }}$r_i > r_j$,\\0~&{\text{ if }}$r_i == r_j$, \\ -1~&{\text{ if }}$r_i < r_j$.\end{cases}

In addition to recognizing the qualification of individual sentences, we believe good explanations should also cover important attributes for each user-item pair. This can be achieved by requiring the learnt attribute representations to be predictive about the ground-truth attributes. As a result, we introduce a logistic regression classifier based on the output representation $\hat{x}_{f_i}$ for each attribute node $f_i$, $p(f_i) = \sigma\big(\langle W^f_o\,, \hat{x}_{f_i}\rangle\big)$, to predict if $f_i$ should appear in the explanation.
We adopt the cross entropy as the loss function of attribute predictions. With the ground-truth label $y_{f_i}$, i.e., those appear in the ground-truth review content for a user-item pair, the loss of attribute predictions is,
\begin{align*}
    L_f = -\sum_{i=1}^{M} y_{f_i}\log p(f_i)
\end{align*}

Combining these two losses, we obtain the objective function as 
\begin{align*}
    L = \lambda L_s+ (1-\lambda)L_f
\end{align*}
where $\lambda$ is a hyper-parameter to control the weight of each loss to the objective. 

% \textbf{Conditional Sentence Prediction.} One drawback of current sentence prediction in Eq.~\ref{eq:xmodel_sent_score} is that it is globally identical across $\mathcal{U}$ and $\mathcal{V}$. Although it is arguably that sentence node representation $\hat{x}_{s_i}$ is implicitly conditioned on the corresponding user and item, in order to emphasize this conditional relationship between user-item pair and sentence, we choose to explicitly modeling this during the sentence prediction stage. Specifically, we adopt Deep \& Cross Network (DCN)~\cite{wang2017deep} which is widely used to capture different level of feature interactions across multiple domains.

\noindent\textbf{Collective Sentence Selection.} To reduce redundancy and increase coverage in the extracted sentences, we should select the sentences that are dissimilar to each other but also highly relevant to the target user-item pair. The scoring function $g(s)$ is trained to optimize the latter, but it alone cannot handle the former, which is a combinatorial optimization problem. 
%Therefore, the optimization objective of selecting $K$ sentences to synthesize an explanation is
%\begin{align*}
%   \text{max} \; \mathcal{O} = relevance(\{S_i\}_{i=1}^{K})- \alpha \cdot repetitiveness(\{S_i\}_{i=1}^{K})
%\end{align*}
%
To find the final $K$ sentences, we adopt Integer Linear Programming (ILP) to solve this optimization problem, which is formulated as follows, 
\begin{align*}
   \text{max} \; & \sum_{i=1}^{|S_g|} x_{s_i} g(s_i)- \alpha \sum_{i\ne j} y_{ij} \textit{sim}(s_i, s_j) \\
s.t. & \sum_{i=1}^{|S_g|} x_{s_i} = K \\
& x_{s_i} \in \{0, 1\}, \; \forall \; i, \\
& y_{ij} \in \{0, 1\}, \; \forall \; \{i, j\}\\
& x_{s_i} + x_{s_j} <= y_{ij}+1 \\
& \sum_{i\ne j} y_{ij} = K*(K-1)
\end{align*}
where $\alpha$ balances between relevance and content redundancy. As the selection of sentences for each user-item pair can be performed independently, this ILP problem can be solved with parallelism.  
\section{Experiments}
In this section, we investigate the effectiveness of our proposed solution \xmodel{} in generating explanations for recommended items. We conducted experiments on two large review-based recommendation datasets and compared our model against a set of state-of-the-art baselines to demonstrate its advantages. In addition, we also performed ablation analysis to study the importance of different components in \xmodel{}.

\subsection{Experiment Setup}

We chose review-based public recommendation datasets Ratebeer \citep{mcauley2012learning} and TripAdvisor \citep{wang2010latent} for our evaluation purpose. Both datasets contain user-provided textual reviews about their opinions towards specific items, including user ID, item ID, review text content and opinion ratings. In the Ratebeer dataset, the ratings fall into the range of $[1, 20]$; and in the TripAdvisor dataset, the rating's range is $[1, 5]$. Since a recommender system would generally recommend items that are attractive to users, we only focused on user-item interactions with positive ratings. In particular, we used Ratebeer reviews with ratings greater than $10$ and TripAdvisor reviews greater than $3$ to construct the corpus for our experiments. Since \xmodel{} focuses on explanation generation, in the experiments we directly used the observed item in each user-item pair as the recommendation to be explained.

\begin{table}[t]
    \centering
    \caption{Summary of the processed datasets. Rb stands for Ratebeer and TA stands for TripAdvisor.}
    \label{tab:xmodel_stats}
    \vspace{-2mm}
    \begin{tabular}{c|ccccc}
    \hline
     & \# Users & \# Items & \# Reviews & \# Sentences  & \# Attributes \\ 
    \hline
    Rb  & 1,664 & 1,487 & 109,746 & 519,353 & 572  \\
    TA  & 4,948 & 4,487 & 159,834 & 560,367 & 503 \\
    \hline
    \end{tabular}
    \vspace{-4mm}
\end{table}

\textbf{Data Pre-Processing.} Review content has been directly used as ground-truth explanations for evaluation in many previous studies~\citep{chen2018neural, wang2018reinforcement}. But as suggested in \citet{ni2019justifying}, a large portion of sentences in a review describes personal subjective experience, like ``\textit{I drank two bottles of this beer}'', which does not provide any information about the reason why the user liked this item, and hence they are not qualified as explanations. In contrast, sentences that serve as explanations should describe the important properties of items to help users make informed decisions, like ``\textit{taste is of bubble gum and some banana.}'' Therefore, we construct the explanation dataset by keeping informative sentences in the experiments. For both datasets, we used the Sentires toolkit \citep{zhang2014users} to extract attribute words from reviews and manually filter out inappropriate ones based on our domain knowledge. Then, for each review, we kept sentences that cover at least one of the selected attribute words as candidate explanations. 

We also filtered inactive users and unpopular items: we kept users who at least have fifteen reviews and items are associated with at least fifteen reviews. We keep the 20,000 most frequent words in our vocabulary and use the ``unk'' token to represent the rest words. The statistics of the processed datasets are reported in Table \ref{tab:xmodel_stats}. We split the dataset into training, validation, and testing dataset according to the ratio $70\%\!:\!15\%\!:\!15\%$.

\textbf{Baselines.} We compare our model with five baselines, covering both generation-based and extraction-based methods, which can produce natural language sentences as explanations :
\begin{itemize}
  \item \textbf{NRT}: Neural Rating and Tips Generation \citep{li2017neural}, a generation-based solution. It is originally proposed for tip (a short sentence summary) generation, but can be seamlessly adapted to generating explanations. It utilizes an RNN-based neural language model to generate explanations.
%   since tips play a similar role as explanations in recommendations.
  \item \textbf{SAER}: Sentiment Aligned Explainable Recommendation \citep{yang2021explanation}. This is another generation-based solution. It focuses specifically on the sentiment alignment between the recommendation score and generated explanations. It implements a sentiment regularizer and a constrained decoding method to enforce the sentiment alignment in the explanations in both training and inference phases.
  \item \textbf{NARRE}: Neural Attentional Regression model with Review-level Explanations \citep{chen2018neural}. It is an extraction-based solution. It learns the usefulness of the existing reviews through attention. It selects the review with highest attention score as the explanation.
  \item \textbf{SEER}: Synthesizing Aspect-Driven Recommendation Explanations from Reviews \citep{le2021synthesizing}. This is another extraction-based solution. It takes a user's sentiment towards item's aspects as input, which can be obtained from explainable recommendation models such as EFM \citep{zhang2014explicit}, and then form an ILP problem to select the $K$ most representative and coherent sentences that fit the user's demand of $K$ aspects.
  \item \textbf{ESCOFILT}: Unsupervised Extractive Summarization-Based Representations for Accurate and Explainable Collaborative Filtering \citep{pugoy2021unsupervised}. This is also an extraction-based solution. It is built on BERT text representation to generate user and item profile by clustering user-side and item-side sentence embeddings using $K$-Means. The $K$ sentences that are the closest to their own cluster centroids are selected to form the explanations.
\end{itemize}

\textbf{Implementation Details.} For both datasets, we first pre-trained 256-dimension Glove embeddings~\citep{pennington2014glove} on the whole vocabulary and fine-tuned BERT model on our dataset using Sentence-BERT~\citep{reimers2019sentence} which were later used to initialize the attribute node and sentence node, respectively. User and item node embedding size $d_{u}$ and $d_{c}$ were both set to 256. We stacked 2 GAT layers, with 4 head in the first layer and 1 head in the second layer. The hidden dimension size in GAT was set to 256. For DCN, we combined a 2-layer cross network with a 2-layer MLP whose hidden size and output size were both set to 128. The sentence relevance score $r_i$ in Eq \eqref{eq-stn-sim} used in pairwise loss was the pre-computed maximum sentence BLEU score between $s_i$ and $\mathcal{S}_{uc}$.

During training, we used a batch size of 16 and applied Adam optimizer~\citep{kingma2014adam} with a learning rate of 2e-4. The $\lambda$ in multi-task loss function was set to 0.5. The model was selected according to its performance on the valid set where we took the top-5 predicted sentences and computed the BLEU score with the ground-truth review. When finally generating the explanations, in order to reduce the computation complexity of the ILP problem, we selected the top-100 predicted sentences for each user-item pair on the test set and use the Gurobi\footnote{https://www.gurobi.com/products/gurobi-optimizer/} solver to select the top-5 most relevant yet non-repetitive sentences. We computed the cosine similarity based on the tf-idf representation of a pair of sentences. $\alpha$ is set to 2.0 according to the performance on the validation set.

\subsection{Quality of the Generated Explanations}
To comprehensively evaluate the quality of the generated explanations at both word-level and attribute-level, we employed different types of metrics, including BLEU-\{1, 2, 4\} and F1 score of ROUGE-\{1,2,L\}, which are used to measure word-level content generation quality, and Precision, Recall and F1 score of mentioned attributes, which are used to measure the attribute-level content generation quality. The results are reported in Table \ref{tab:xmodel_exp_eval} and Table \ref{tab:xmodel_exp_eval_attr} respectively.

\begin{table}[t]
    \caption{Comparison of word-level explanation quality by different models on Ratebeer and TripAdvisor.}
    \centering
    \label{tab:xmodel_exp_eval}
    \vspace{-2mm}
    \begin{tabular}{c|ccc|ccc}
        \hline
        \multirow{2}{*}{Model} & \multicolumn{3}{|c|}{BLEU(\%)} & \multicolumn{3}{|c}{ROUGE(\%)}  \\
        & 1 & 2 & 4 & 1 & 2 & L  \\
        \hline
        \multicolumn{7}{c}{Ratebeer} \\
        \hline
        NRT & 27.03 & 11.97  & 2.50 & 27.16 & 4.83 & 24.63  \\
        SAER & 28.40 & 12.68 & 2.66 & 27.59 & 4.92 & 25.29 \\
        NARRE & 24.67 & 9.09 & 1.41 & 22.13 & 2.79 & 20.04 \\
        SEER & 14.24 & 5.77 & 1.03 & 20.18 & 2.74 & 21.08 \\
        %SEER (15 aspects) & 29.84 & 13.30 & 2.60 & 26.09 & 3.98 & 27.35 \\
        ESCOFILT & 26.36 & 12.27 & 3.55 & 27.55 & 5.58 & 24.62 \\
        % ESCOFILT (only item) & 24.24 & 9.18 & 1.35 & 24.67 & 3.28 & 21.85 \\
        \xmodel{} & \textbf{36.59} & \textbf{19.14} & \textbf{6.15} &  \textbf{33.03} & \textbf{8.34} & \textbf{29.84} \\
        \hline
        \multicolumn{7}{c}{TripAdvisor} \\
        \hline
        NRT & 20.30 & 8.31 & 1.70 & 20.25 & 2.92 & 18.86  \\
        SAER & 20.94 & 8.80 & 1.90 & 20.67 & 3.23 & 19.23 \\
        NARRE & 22.22 & 8.59 & 1.67 & 21.92 & 2.85 & 19.52 \\
        SEER & 21.84 & 9.00 & 1.89 & 21.77 & 3.13 & 20.44 \\
        %SEER (10 aspects) & 24.70 & 11.05 & 2.53 & 24.64 & 4.12 & 23.37 \\
        ESCOFILT & 22.81 & 9.40 & 2.39 & 22.94 & 3.41 & 20.42 \\
        % ESCOFILT (only item) & 22.96 & 9.36 & 1.96 & 23.69 & 4.18 & 20.81 \\
        \xmodel{} & \textbf{28.78} & \textbf{13.39} & \textbf{4.38} & \textbf{24.99} & \textbf{4.90} & \textbf{22.33} \\
        \hline
    \end{tabular}
\end{table}

\textbf{Word-level Content Generation Quality.} \xmodel{} outperformed baselines under every BLEU and ROUGE metric on both datasets. As BLEU is a precision-based metric, a larger BLEU achieved by a model suggests that a larger portion of content in sentences selected by the model is relevant to the ground-truth sentences. In other words, more content in sentences selected by \xmodel{} can reflect/predict the users' opinions towards the recommended items. 
%The comparison between \xmodel{} and NRT and SEAR shows that the graph structure employed in \xmodel{} better helps the model fuse users' preferences and items' properties to  sentences than the sequential structure modeled by RNN does. 
RNN-based generative methods such as NRT and NARRE suffer from generating short and generic content, such as ``\textit{the staff was very friendly and helpful.}'' Although such generic content has a larger chance to match with the ground-truth sentences, the brevity penalty in BLEU penalizes methods focusing too much on such less information sentences. The comparison of \xmodel{} against NARRE shows that though both methods extract sentences from an existing corpus, the graph structure employed in \xmodel{} can better recognize sentences matching users' criteria of items than the attention structure used in NARRE. 
%Besides, compared to NARRE, \xmodel{} extracted sentences from a finer-grained level rather than the whole review, which results in higher flexibility on choosing sentences. 
Both \xmodel{} and SEER select the final $K$ sentences using ILP. The comparison between them shows that \xmodel{}'s learnt sentence relevance score better reflects the utility of a sentence as an explanation than directly modeling it from pre-computed user sentiment scores over item aspects. 
%The comparison between \xmodel{} and ESCOFILT shows that \xmodel{}'s graph structure can capture user's preference on item's aspects and generate personalized explanations rather than semantically summarization of the item's profile. Although ESCOFILT's generated summarization may arguably cover more aspects about the target item, it's not always suitable from the user's perspective since she may only concentrate on a subset aspects of the item which explain the dropping on BLEU score comparing it with \xmodel{}. 
ESCOFILT is enforced to cover $K$ different aspects of an item for its explanations to all users, while \xmodel{} learns to predict what the most important sentences are for each user-item pair. This hard requirement in ESCOFILT makes it one of the most competitive baselines, but its lack of flexibility also leaves it behind \xmodel{}. 
It is noteworthy that \xmodel{} achieves much larger BLEU-4 than all baselines. It suggests the explanations generated by \xmodel{} not only have more overlaps with the ground-truth, but also cover longer segments in the ground-truth; and therefore its content is more coherent as a whole. 
%This suggests the portion of 4-grams in sentences selected by \xmodel{} overlapping 4-grams in ground-truth sentences is much larger than that in sentences selected by other models. 

On the other hand, as ROUGE is a recall-based metric, a higher ROUGE score achieved by a model suggests that more content in ground-truth sentences are included in the sentences selected by the model. 
%Thus, the superior performance achieved by \xmodel{} suggests that \xmodel{} can recognize more content relevant to users about recommended items. 
This further demonstrates the effectiveness of \xmodel{} in recognizing the relevance of candidate sentences to a user-item pair. In particular, \xmodel{} achieved much higher ROUGE-L than all baselines. This indicates that \xmodel{} is more successful in identifying sentences with long consecutive spans that match users' comments about items. The graph structure that fuses heterogeneous information from users, items, attributes and sentences contributes the most to \xmodel{}'s better extraction performance, which will be validated in our ablation analysis in Section \ref{sec-ablation}.

\begin{table}[h]
    \caption{Comparison of attribute-level explanation quality by different models on Ratebeer and TripAdvisor.}
    \centering
    \label{tab:xmodel_exp_eval_attr}
    % \vspace{-2mm}
    \begin{tabular}{c|ccc|ccc}
        \hline
        \multirow{2}{*}{Model} & \multicolumn{6}{|c}{Attribute Prediction(\%)}  \\
        & P & R & F1 & P & R & F1  \\
        \hline
        Dataset & \multicolumn{3}{|c|}{Ratebeer} & \multicolumn{3}{|c}{TripAdvisor}\\
        \hline
        NRT   & 29.27 & 23.72 & 24.76 & 22.82 & 19.22 & 17.82 \\
        SAER  & 28.82 & 23.75 & 24.57 & \textbf{22.95} & 19.23 & 18.02 \\
        NARRE & 20.19 & 24.38 & 20.08 & 17.32 & 24.95 & 18.21 \\
        SEER  & \textbf{31.45} & 22.57 & 24.17 & 22.66 & 26.15 & 21.95 \\
        ESCOFILT  & 21.32 & 32.99 & 24.39 & 16.92 & 28.21 & 19.26 \\
        \xmodel{} & 30.60 & \textbf{40.02} & \textbf{32.74} & 20.92 & \textbf{33.68} & \textbf{23.46} \\
        \hline
    \end{tabular}
    % \vspace{-2mm}
\end{table}

\textbf{Attribute-level Content Generation Quality.} To better understand whether the explanations generated by \xmodel{} cover more important information about the target items' attributes, we also evaluated the Precision, Recall and F1 score of the attributes contained in the synthesized explanation with respect to the corresponding ground-truth. As we can observe in Table \ref{tab:xmodel_exp_eval_attr}, \xmodel{} outperformed baselines with a large margin in recall and comparable precision on both datasets. This strongly suggests \xmodel{} can better recognize the relevance of candidate sentences with respect to the important item attributes, and maximize the coverage of those attributes. On the other hand, it confirms that \xmodel{}'s encouraging performance on BLEU and ROUGE is not simply because of its extractive nature, but also due to its ability to more accurately identify and cover important target attributes.

% \xmodel{} outperforms baselines under Recall and F1 score of the attribute prediction on both datasets. Compared to generation-based methods including NRT and SAER, \xmodel{} obtains similar precision score on both datasets, but significantly higher recall and F1. This indicates that although generation-based methods output explanations with precise attributes, they contain fewer number of attributes (mainly generic ones) compared to \xmodel{}. For extraction-based methods, SEER possesses larger precision score but smaller recall score which is reasonable as it's demanded to generate sentences about the top-5 most popular aspects for each user-item pair. ESCOFILT, on the other hand, possesses large recall score but smaller precision score which confirms our argument that ESCOFILT's generated summarization doesn't always contains the attributes that the user concentrated on.

\subsection{Ablation Analysis}
\label{sec-ablation}
We include four variants of our solution to study the contribution of each component to the performance of \xmodel{}:
\begin{itemize}[leftmargin=*]
    \item \textit{$\neg$ GAT.} This variant excludes the graph neural network model and only preserves the user, item, attribute and sentence information. For each sentence $\mathcal{S}_{uc}$, we concatenate its embedding with user $u$, item $c$ and mean-pooling of $f_{\mathcal{S}_{uc}}$, where $f_{\mathcal{S}_{uc}}$ represents the attributes in sentence $\mathcal{S}_{uc}$. Then the concatenated embedding is directly fed into DCN to predict the relevance score of sentence $\mathcal{S}_{uc}$. The comparison between this variant and \xmodel{} indicates the necessity of modeling the user, item, attributes and sentences into a heterogeneous graph to learn their intermediate relationships.
    \item \textit{$\neg$ BERT.} This variant replaces BERT with the average word embeddings for sentence representations. The comparison between this variant and \xmodel{} demonstrates the importance of using a pre-trained language model to encode sentences as input sentence node representations.
    \item \textit{$\neg$ DCN.} This variant replaces DCN in \xmodel{} with a single linear layer. The comparison between this variant and \xmodel{} presents the effectiveness of including direct feature-level interaction among user, item and sentences when learning the final sentence representations.
    \item \textit{$\neg$ ILP.} This variant replaces the ILP-based sentence selection strategy with a vanilla strategy which selects sentences solely based on the predicted probabilities of sentences in a descending order. The comparison between this variant and \xmodel{} shows the utility of considering both sentence relevance and redundancy when selecting sentences as explanations.
\end{itemize}

\begin{table}[h]
    \caption{Ablation analysis on Ratebeer dataset.}
    \label{tab:xmodel_ablation}
    \centering
    \begin{tabular}{l|ccc|ccc}
        \hline
        \multirow{2}{*}{Model} & \multicolumn{3}{|c|}{BLEU (\%)} & \multicolumn{3}{|c}{ROUGE (\%)}  \\
        & 1 & 2 & 4 & 1 & 2 & L  \\
        \hline
        \multicolumn{7}{c}{Ratebeer} \\
        \hline
        \xmodel{} & \textbf{36.59} & \textbf{19.14} & \textbf{6.15} & \textbf{33.03} & \textbf{8.34} & \textbf{29.84} \\
        $\neg$ GAT & 33.95 & 17.74 & 5.88 & 31.25 & 8.14 & 28.23 \\
        $\neg$ BERT & 33.26 & 16.59 & 4.96 & 32.16 & 7.40 & 28.88 \\
        $\neg$ DCN & 35.13 & 17.63 & 5.35 & 31.87 & 7.31 & 29.08 \\
        $\neg$ ILP & 27.73 & 14.79 & 5.75 & 24.59 & 7.46 & 26.01 \\
        \hline
    \end{tabular}
\end{table}

The results of our ablation analysis about \xmodel{} are reported in Tabel~\ref{tab:xmodel_ablation}. The same as Table~\ref{tab:xmodel_exp_eval}, we report BLEU-\{1, 2, 4\} and F1 score of ROUGE-\{1,2,L\}. All the variants performed worse than \xmodel{}. The most important component turns out to be ILP, which is expected. As the learnt sentence embedding in \xmodel{} reflects the relevance of a sentence to a given user-item pair independently from other sentences, simply counting on this sentence representation for selecting sentences can hardly avoid repetitions. When we looked into the output of $\neg$ ILP, most of its top ranked sentences are very similar to each other and therefore can hardly cover a comprehensive set of aspects of the target item. 
%plays a significant role on extracting sentences that are both relevance and less repetitive which is crucial to generate perceivable and trustworthy explanations.
The next most important component is the heterogeneous graph structure, which helps \xmodel{} capture the complex relationship between a user's preference and item's aspects, which cannot be tackled by directly performing feature crossing. 
In addition, using BERT to obtain sentence representations enables \xmodel{} to learn a better sentence representation. 
This analysis suggested that DCN introduced least impact on \xmodel{}'s final performance; but combining DCN with GAT can still boost the model's performance. This indicates feature-level interactions in the embedding space provide a complementary view for the final sentence representation and selection. 

% All these variants perform worse than \xmodel{}, which suggests that modeling this graph structure indeed help the model to learn the relationship between user's preference and item's aspects which cannot be easily tackled by directly performing feature crossing. In addition, using BERT to obtain sentence representations enables \xmodel{} to make use of the co-occurrence of feature-sentence to recognize relevant sentences. Moreover, although the variant without DCN can already reach a promising result, combining DCN with GAT can further boost the performance which indicates that explicitly emphasizing user and item by feature crossing during the sentence prediction stage helps the model to choose sentences that better reveal the user's preference and item's attributes. Finally, ILP plays a significant role on extracting sentences that are both relevance and less repetitive which is crucial to build a perceivable and trustworthy explainable recommendation system.

% \subsection{Hyperparameter}

\subsection{Case Study}

We present two group of example explanations produced by \xmodel{} and other baselines in Table \ref{table:xmodel_example}. The ground-truth explanations are also included for reference. We manually labeled the overlapping attributes in the generated sentences from the 3 methods. From the table, we can clearly observe that the extracted sentences by \xmodel{} are much more relevant to the ground-truth explanations. The features in extracted sentences match those in ground-truth explanations, especially those uncommon attributes such as ``roasty chocolate'' in the first example and ``hot tub'' in the second example. With the explicitly mentioning of features, the explanations generated by \xmodel{} can help users make more informed decisions on which item better suits their preferences and needs.

\begin{table*}[h]
    \caption{Example explanations produced by different models on Ratebeer and TripAdvisor.}
    \vspace{-2mm}
    \label{table:xmodel_example}
    \begin{tabular}{|cp{12cm}|}
        \hline
        Model & Explanation \\
        \hline
        \multicolumn{2}{|c|}{Ratebeer} \\
        \hline
        Ground-Truth & Roasty, chocolate malts paired with chinook hops. Very smooth sipper with a nice balance between sweet, smooth malt, and tangy hops. \\
        NARRE & Medium carbonation and body; with a very nice creamy and slightly slick mouthfeel. \\
        SAER & A dark orange color with a medium - sized white head. \\
        \xmodel{} & Tastes of \textbf{hops} and \textbf{roasty chocolate}.Taste is very \textbf{smooth} with bitter and herbal \textbf{hops} with creamy toasted \textbf{malts}.  \\
        \hline
        \multicolumn{2}{|c|}{TripAdvisor} \\
        \hline
        Ground-Truth & The rooms are the perfect size and have everything you would need. The outdoor hot tub is also fantastic for your aching legs after a full day of skiing. \\
        NARRE &  The hotel is very centrally located so we were able to walk which was really nice. \\
        SAER & Great hotel for breakfast, and the staff are very friendly. \textbf{Room} is clean and spacious. \\
        \xmodel{} & The \textbf{hot tub} is great and a good place to meet people. Comfortable beds, nice large bathrooms. maybe the cleanest motel \textbf{room} I've ever been in. \\
        \hline
    \end{tabular}
    \vspace{-2mm}
\end{table*}

\section{Conclusion and Future Work}
In this paper, we present a graph neural network based extractive solution for explaining a system's recommended items to its users. It integrates heterogeneous information about user, item, item attributes and candidate sentences to evaluate the relevance of a sentence with respect to a particular user-item pair. Item attributes are introduced as the intermediary to address sparsity in observations at the user-item level, and for the same purpose pre-trained language models are used to encode item attributes and sentences for the model learning. Finally, to optimize the trade-off among individual sentence relevance, overall attribute coverage and content redundancy, we solve an integer linear programming problem to make the final selection of sentences for a user-item pair. Extensive experiment comparisons against a set of state-of-the-art explanation methods demonstrate the advantages of our solution in providing high-quality explanation content.

We should note extraction-based explanation methods still have their intrinsic limitations: their input is restricted to an item's existing reviews; for items with limited exposure, e.g., a new item, such solutions (including ours) cannot provide any informative explanations. Our current attempt to address this limitation was to incorporate the same user's historical reviews about items from the same category. Leveraging generative solutions to synthesize explanations could be a potential choice in such situations. In addition, currently the scoring function's weights in our ILP were manually set. Learning-based methods can be introduced to optimize it for better performance in our future work.

% Acknowledgements should go at the end, before appendices and references

\acks{This work is supported by the National Science Foundation under grant IIS-1553568, IIS-1718216 and IIS-2007492. }

% Manual newpage inserted to improve layout of sample file - not
% needed in general before appendices/bibliography.

% Note: in this sample, the section number is hard-coded in. Following
% proper LaTeX conventions, it should properly be coded as a reference:

%In this appendix we prove the following theorem from
%Section~\ref{sec:textree-generalization}:

\newpage
\bibliography{reference}

\end{document}